\documentclass[prl,aps,amssymb,amsmath,superscriptaddress,twocolumn,showpacs]{revtex4-1}

\usepackage{graphicx}

\newcommand{\disp}{\displaystyle}
\newcommand{\tx}{\textstyle}
\newcommand{\lf}{\left}
\newcommand{\ri}{\right}
\newcommand{\diff}{\mathrm{d}}
\newcommand{\spinv}{{\bf S}}
\newcommand{\bra}[1]{\lf\langle #1\ri|}
\newcommand{\ket}[1]{\lf|#1 \ri\rangle}

\newcommand{\D}{\mathcal{D}}
\newcommand{\e}{\mathbf{e}}
\renewcommand{\epsilon}{\varepsilon}

\renewcommand{\phi}{\varphi}

\newcommand{\Z}{\mathbb{Z}}

\DeclareMathOperator{\Tr}{Tr}
\DeclareMathOperator{\tr}{Tr}

\def\be{\begin{equation}}
\def\ee{\end{equation}}
\def\bea{\begin{eqnarray}}
\def\eea{\end{eqnarray}}

\def\xx{{\bf x}}
\def\yy{{\bf y}}
\def\zz{{\bf z}}
\def\zv{\vec{z}}

\def\pp{{\bf p}}

\def\OO{\mathcal{O}}

 \let\b=\beta         \let\d=\delta    
\let\e=\varepsilon
        \let\k=\kappa

\let\r=\rho
            
\let\c=\chi
        
\let\G=\Gamma \let\D=\Delta       \let\L=\Lambda

\begin{document}

\title{Validity of spin wave theory for the quantum Heisenberg model}

\author{Michele Correggi}
\author{Alessandro Giuliani}
\affiliation{Dipartimento di Matematica e Fisica, Universit\`a di Roma
Tre, L.go S. L. Murialdo 1, 00146 Roma, Italy}

\author{Robert Seiringer}
\affiliation{Institute of Science and Technology Austria (IST Austria), Am
Campus 1, 3400 Klosterneuburg, Austria}

\begin{abstract}
Spin wave theory is a key ingredient in our comprehension of quantum spin systems,
and  is used successfully for understanding a wide range of magnetic phenomena, including  magnon condensation and 
stability of patterns in dipolar systems. Nevertheless, several decades of research failed to establish the validity of spin wave theory rigorously, 
even for the simplest models of quantum spins.
A rigorous justification of the method for the three-dimensional quantum Heisenberg ferromagnet at low temperatures is presented here.
We derive sharp bounds on its free energy by combining  a bosonic formulation 
of the model introduced by Holstein and Primakoff 
with probabilistic estimates and operator inequalities. 
\end{abstract}

\pacs{05.30.-d, 
75.10.Jm, 75.30.Ds}

\maketitle


The quantum Heisenberg ferromagnet (QHF) is one of the simplest models used to describe the phenomenon of spontaneous
breaking of a continuous symmetry.  The understanding of its low-temperature properties is mostly based
on spin-wave theory, which predicts a phase transition in three or more dimensions, and the $T^{3/2}$ Bloch law for 
the magnetization, whose experimental verification dates back to the 1960s \cite{ACP}. More recently, spin-wave theory was successfully used to 
investigate Bose-Einstein condensates of magnons in ferromagnetic nanostructures \cite{DBW,Detal,LSP} and in magnetic insulators \cite{Netal, GRT}, as well as the stability of patterns in dipolar thin films \cite{DMW}.
Despite its simplicity and its 
reliable predictions, a rigorous control of the spin-wave expansion remains to date a challenge. 
In the case of an underlying abelian symmetry, a number of rigorous results are available, based on reflection positivity \cite{FSS, DLS, BFLLS}, or cluster expansion combined with a vortex loop representation \cite{FS, KK}. The non-abelian case is trickier, 
and the few results available are mostly based on reflection positivity: see \cite{FSS} for the classical Heisenberg 
and \cite{DLS} for the quantum Heisenberg anti-ferromagnet. 

In this letter, we present the key ingredients in a rigorous proof of the validity of the spin wave approximation at the level 
of the first non-trivial contribution to the free energy of the QHF in three dimensions at low temperatures. 
It is the first rigorous confirmation of the predictions of Bloch and Holstein-Primakoff. It comes more than 80 years 
after  the original formulation of spin-wave theory, and after more than 40 years of efforts of the mathematical physics 
community. While our 
method is not capable yet to control the spontaneous magnetization, it introduces new ideas in the field
by deriving two novel rigorous  inequalities, on the low-energy spectrum of the quantum spin model, as
well as on the two-point function. These estimates allows us to rigorously reduce the 
many-body problem to a two-body one, which can be studied by probabilistic techniques. 
In comparison with methods based on reflection positivity, our method is robust:
we do not expect that the results depend crucially on the underlying lattice structure, or on the nearest neighbor nature of the interaction. Still, in order to make our ideas as transparent as possible, we stick here to the simplest version of the model: we consider the Hamiltonian
\begin{equation}\label{heisenberg ham 1}
H_\Lambda =  \sum_{\langle \xx,\yy \rangle \subset \Lambda} \lf(S^2 - {\bf S}_{\bf x} \cdot {\bf S}_{\bf y} \ri)\,,
\end{equation}
where $\Lambda\subset \Z^3$ is a cube, the sum is over all (unordered) nearest neighbor pairs $\langle \xx,\yy\rangle$ in $\Lambda$, and ${\bf  S_{x}} $ is a spin $ S $ operator with components $ {\bf S}_{\bf x} = (S^1_{\bf x}, S^2_{\bf x}, S^3_{\bf x}) $. The constant $S^2$ is chosen to normalize the ground state energy of $H_\Lambda$ to zero. 
We denote the specific free energy in $\L$ by 
\be f(S,\b,\L)=-\frac1{\b|\L|}\ln \Tr e^{-\b H_\L}\;,\nonumber\ee
and by $f(S,\b)$ its value in the thermodynamic limit.

\medskip

{\bf Main result:}	For any $S\geq 1/2$, we have
	\begin{equation}\label{def:c0}
 f(S,\beta) \simeq  \frac1\b\int \ln \left( 1- e^{-\b S\e(\pp)}\right) \frac{\diff {\bf p}}{(2\pi)^{3}} 
\end{equation}
 to leading order in $\b$ as $\b\to\infty$, where $\e(\pp)=2\sum_{i=1}^3(1-\cos p_i)$.

\medskip

The right side of \eqref{def:c0} is the free energy of a non-interacting lattice Bose gas with nearest neighbor hopping of amplitude $S$, and is predicted by spin wave theory.
Asymptotically, it equals $C_0S^{-3/2}\b^{-5/2}$, with $C_0=-0.030...$
The proof is based on rigorous upper and lower bounds. Until now, at finite $S$ not even a sharp upper bound was known. Two non-optimal upper bounds were derived in \cite{CS,T}. 
Sharp upper and lower bounds in 
a suitable large-$S$ limit were derived in \cite{CG}.

An important consequence of our proof is an instance of quasi long-range order: with $\langle\cdot\rangle_\b$ a translation invariant Gibbs state at inverse temperature $\beta$,
\be \langle S^2-{\bf S}_\xx\cdot{\bf S}_\yy\rangle_\b\le \tfrac{27}8|\xx-\yy|^2e(S,\b)\;,\label{eq.3}\ee
where $e(S,\b)=\partial_\b (\b f(S,\b))$ is the energy per site. Our main result says that 
$e(S,\b)\simeq -\tfrac32C_0S^{-3/2}\b^{-5/2}$ for large $\b$. Therefore, 
Eq.\,\eqref{eq.3} implies that order persists up to length scales of the order $\b^{5/4}$, i.e., $\langle {\bf S}_\xx\cdot{\bf S}_\yy\rangle_\b$ is bounded away from zero
as long as $|\xx-\yy|\le ({\rm const.})\b^{5/4}$. Spin wave theory predicts equality in \eqref{eq.3} without the factor $\tfrac{27}8$, asymptotically  for $|\xx-\yy|\ll \sqrt\b$. 
Of course, one expects infinite range order at low temperatures, but in absence of a proof Eq.\,\eqref{eq.3} is the best result to
date.

In the following we spell out the proof of \eqref{def:c0} for $S=1/2$ only, and refer to \cite{CGS} for the general case and additional details. For short, we denote $f(1/2,\b)$ by $f(\b)$.

{\it Bosonic representation.} It is well known that the Heisenberg Hamiltonian can  be rewritten in terms of bosonic creation and annihilation operators \cite{HP}. 
The spin Hilbert space is mapped onto the bosonic Fock space with the additional constraint that there is at most one particle per site. 
For any $ {\bf x} \in \Lambda $ we set
\begin{equation}\nonumber
	S_{\bf x}^+ =   a^\dagger_{\bf x} (1 - n_{\bf x} ),		\quad S_{\bf x}^{-} = (1  - n_{\bf x} )a_{\bf x},	\quad	S_{\bf x}^3 =  n_{\bf x}-\tx\frac{1}{2},
\end{equation}
where $a^{\dagger}_{\xx}, a_{\xx}$ are bosonic creation and annihilation operators,  $n_\xx=a^\dagger_\xx a_\xx$ and $S^\pm = S^1 \pm i S^2$.  The Hamiltonian $H_\Lambda$ in (\ref{heisenberg ham 1})  can be expressed as
\be  H_\Lambda  =  \tx\frac12 P \disp\sum_{\langle \xx,\yy\rangle\subset \Lambda} 
\Big[(   a^\dagger_\xx -a^\dagger_\yy)(a_\xx-a_\yy)- 2 n_\xx n_\yy\Big] P
\label{s.2}\ee
which we write as $H_\L=PTP-K$, where $P$ is a projection that enforces the hard-core constraint and $K$ is the nearest neighbor 
density-density interaction.

{\it Upper bound.} We localize the system into Dirichlet boxes of side $\ell$, to be optimized over:
we pave $\L$ using cubes $B$ of side $\ell$ plus 
one-site-thick corridors between them. Since $H_\L\le \sum_{B\subset\L} H^D_B$, where $H^D_B$ is the Hamiltonian with $S^3_x=-1/2$ (i.e., Dirichlet)
boundary conditions on $B$, $f(\beta,\Lambda)$ is bounded above by $( 1 + \ell^{-1} )^{-3} f^D(\beta,B)$, 
with $f^D(\beta,B)= - \frac 1{\beta |\Lambda|} \ln \Tr e^{-\beta H_B^D}$.
In each box $B$, we use the Gibbs variational principle:
\begin{equation}
f^D(\beta,B)=\frac{1}{\ell^3} \min_{\G} \Big[\tr H^D_{B} \Gamma  + \frac{1}{\beta} \tr  \Gamma \ln \Gamma\Big]\nonumber
\end{equation}
where one minimizes over normalized density matrices. In order to get an upper bound on the right side, we use as trial state $\G_0=P e^{-\b T^D} P/$(normalization), where 
$T^D$ is the hopping term with Dirichlet boundary conditions,  
$P=\prod_{\bf x} P_{\bf x}$ and $P_{\bf x}$ projects onto $n_\xx\le 1$. The key observation is  that one can get rid of the projectors by exploiting the simple inequality $ 1- P \leq \sum_\xx (1- P_{\bf x}) \leq \frac 12\sum  n_{\bf x} (n_{\bf x}-1) $. Wick's rule for Gaussian states can then be applied to compute the error due to the hard-core constraint. If $\sqrt\b\ll \ell \ll {\b} $, the result is that
\be f(\beta) \leq \frac1{\b\ell^3}\sum_{\pp}\ln(1-e^{-\frac12\b\e(\pp)})+\OO(\b^{-3})+\OO(\ell^3\b^{-11/2})\nonumber\ee
where the sum runs over the Dirichlet wave vectors
in the box $B$. 
The error for replacing the discrete Riemann sum by the corresponding integral is $\OO(\ell^{-1}\b^{-2})$. The optimal choice of  $\ell$ is then $\ell\propto\b^{7/8}$, so that
$$ f(\beta) \leq
 C_0 \lf( \tx\frac12 \ri)^{-3/2} \beta^{-5/2} \left(1 - \mathcal O( \beta^{-3/8}) \right).$$
 
{\it Lower bound.} The proof is divided into three steps: localization and preliminary lower bound; restriction of the trace to the low-energy sector; estimate of the interaction in the low-energy sector.

{\it Step 1.}
We localize the system into boxes $B$ of side $\ell$, to be optimized over: dropping the positive interaction between different boxes we get
\be f(\b,\L)\ge f(\b,B).\label{8}\ee
We now derive a preliminary bound on the free energy of the form $f(\b,B)\ge -({\rm const.}) \b^{-5/2}(\ln\b)^{5/2}$, which relies on the following key lemma. It quantifies the 
minimal energy of states with total spin smaller than the maximum. Apart from the prefactor, it verifies the prediction of spin wave theory. 
We denote by $ S_T $ the quantum number associated to the total spin operator $ {\bf S}_T = \sum_{\xx\in B} {\bf S}_{\bf x} $, i.e., $ |{\bf S}_T|^2 = S_T(S_T+1) $.

\medskip

{\it Lemma 1}. $H_B \ge ({\rm const.})\ell^{-2} \lf(\tx\frac12 \ell^3 - S_T \ri)$.
\medskip

{\it Proof.} For distinct sites $ {\bf x}, {\bf y}, {\bf z} $ we first prove that $$ \left( \tx\frac14  - \spinv_{{\bf x}} \cdot \spinv_{{\bf y}} \right) + \left( \tx\frac14 - \spinv_{{\bf y}} \cdot\spinv_{{\bf z}} \right) \geq \tx\frac{1}{2} \left( \tx\frac14 - \spinv_{{\bf x}} \cdot \spinv_{{\bf z}} \right), $$
which is equivalent to $ \frac{1}{4} |\spinv_{\bf x} + \spinv_{\bf z}|^2 - \spinv_{\bf y} \cdot \lf( \spinv_{\bf x} + \spinv_{\bf z} \ri) \geq 0 $. The operator $ |\spinv_{\bf x} + \spinv_{\bf z}|^2 $ has only eigenvalues 0 and 2: in the first case the inequality is trivially true, while in the second it is sufficient to observe that $ \spinv_{\bf y} \cdot \lf( \spinv_{\bf x} + \spinv_{\bf z} \ri) $ has maximal eigenvalue $1/2$. By repeatedly applying the above inequality, 
one finds that for any $ n +1 $ distinct sites 
\be  \sum_{j}  \left( \tfrac{1}{4}  - \spinv_{{\bf x}_{j}} \cdot \spinv_{{\bf x}_{j+1}} \right)  \geq \tfrac{1}{2n}  
 \left( \tfrac14 - \spinv_{{\bf x}_{1}} \cdot \spinv_{{\bf x}_{n+1}} \right).\label{qq}\ee For any given pair of sites $ {\bf x} $ and $ {\bf y} $, we pick the shortest lattice path connecting the two points that stays as close as possible to the straight line from ${\bf x} $ to $ {\bf y}$ (call it $\mathcal C_{\xx,\yy}$), and estimate
$$ \sum_{{\bf x} \neq {\bf y}} \left(  \tx\frac14 - \spinv_{\bf x} \cdot \spinv_{\bf y} \right) 	\\	\leq 6 \ell  \sum_{|{\bf x}-{\bf y}|=1}  \left(  \tx\frac14 - \spinv_{\bf x} \cdot \spinv_{\bf y} \right)  N_{{\bf x},{\bf y}},$$
where $ N_{{\bf x},{\bf y}} $ denotes the number of  paths among all the $\mathcal{C}_{\zz,\zz'}$, $\zz,\zz'\in B$, that contain the step $\xx\to \yy$. Since the left side equals $\tfrac12 \ell^3 ( \tfrac12 \ell^3 + 1) - S_T(S_T + 1)$ and $ N_{{\bf x},{\bf y}} \leq ({\rm const.}) \ell^4 $, this immediately implies the desired result. 
$\hfill\Box $

\medskip

The strategy of the proof above can also be used to infer \eqref{eq.3}: using \eqref{qq}, we can bound 
the left side of \eqref{eq.3} by twice the square of the number of bonds needed for reaching $\xx$ from $\yy$ on the lattice, times the 
bond energy. This leads to the right side of \eqref{eq.3}, with the factor $\tfrac{27}{8}$ replaced by $6$. A closer inspection yields the stated constant.

Lemma 1 immediately implies an upper bound on the partition function: the number 
of states with total spin $S_T=\frac12\ell^3-N$ is 
$$(\ell^3-2N+1)\left({\ell^3 \choose N}-{\ell^3 \choose N-1}\right)$$
and is smaller than $(\ell^3+1){\ell^3 \choose N}$, hence 
$$ {\rm Tr}(e^{-\b H_B}) \le (\ell^3+1)( 1+  e^{- ({\rm const.}) \beta \ell^{-2}})^{\ell^3}.$$ 
Picking $ \ell \propto \sqrt{\b/\ln\b} $ and using \eqref{8},
we find
\be
	f(\b,\L)\ge -({\rm const.}) \b^{-5/2}(\ln\b)^{5/2},
\label{12}\ee
which is valid in domains $\L$ of side larger than $\sqrt{\b/\ln\b}$.

{\it Step 2.} From now on we choose boxes of side 
$\ell=\b^{1/2+\e}$ with $\e>0$ a small parameter to be optimized in the following.  
We use \eqref{12} to cut off the ``high-energy" sector: if $\c(condition)$ is the characteristic function of the 
set where $condition$ is verified,
$$\tr \c(H_B\ge E_0) e^{-\beta H_{B}} \leq  e^{-\beta E_0/2} e^{-\frac{\beta}2 \ell^3 f(\beta/2,B)} \,.$$
By \eqref{12}, this is smaller than $1$ if $E_0=C \ell^3\b^{-5/2}(\ln\b)^{\frac52}$, for a suitable $C>0$.

We are left with the trace restricted to $H_B\le E_0$, which we compute in sectors at fixed $S_T$ and $S_T^3$. Because of 
$SU(2)$ invariance, the result is independent of $S_T^3$, which we can thus take to be minimal, i.e., $S_T^3=-S_T$.
The degeneracy factor $2S_T+1$ can be bounded by $\ell^3+1$
and, therefore,
\be\tr \c(H_B\le E_0)e^{-\b H_B}\le (\ell^3+1)\tr_{E_0}e^{-\b H_B },\label{13}\ee
where $\tr_{E_0}$ indicates the trace in the subspace $\mathcal H_{E_0}$ with $H_B\le E_0$ and $S^3_T=-S_T$. On this subspace we pass to the bosonic representation
\eqref{s.2}. In this representation the total number of particles equals $N=\frac12\ell^3-S_T$, which by Lemma 1 is bounded above by (const.)$\ell^2H_B$.
It is worth stressing that, by fixing $\ell=\b^{1/2+\e}$, the energy cut-off is $E_0\simeq \ell^{-2+\mathcal O(\e)}$ and hence 
the particle number in $\mathcal  H_{E_0}$ 
is smaller than $\ell^{\mathcal O(\e)}$.

By means of the Peierls-Bogoliubov inequality, 
\be  \tr_{E_0}e^{-\b H_B }\le \tr_{E_0}e^{-\b T} e^{\b\langle K\rangle_{E_0}}\label{1.pb}\ee
where $\langle K\rangle_{E_0}=\tr_{E_0}Ke^{-\b T}/\tr_{E_0}e^{-\b T}$.  We
are left with deriving an upper bound on $ \langle K \rangle_{E_0} $.

{\it Step 3.} 
In order to bound the mean value of the interaction, we first estimate
$$ \langle E| K|E\rangle = \sum_{\langle {\bf x}, {\bf y} \rangle\subset B} \bra{E} n_{\bf x} n_{\bf y} \ket{E} \leq 3 \ell^3\max_{\xx,\yy}\r_E(\xx,\yy),$$
where $|E\rangle$ is an eigenstate of $H_B$ in $\mathcal H_{E_0}$ with energy $E$ and $\r_E(\xx,\yy)$ is the diagonal part of the corresponding  two-particle density matrix. 
The key estimate that we use is the following.

\medskip

{\it Lemma 2.} $\max_{\xx,\yy}\r_E(\xx,\yy)\le ({\rm const.})E^5\ell^4$.
\medskip

Using this and recalling that $\ell=\b^{1/2+\e}$ and $E\le E_0\simeq \ell^{-2+\OO(\e)}$, 
we conclude that $ \langle K\rangle_{E_0} \le ({\rm const.})\ell^{-3+\OO(\e)}$. We now plug 
this 
bound into \eqref{1.pb}. The term $\tr_{E_0}e^{-\b T}$ gives rise to the 
(Riemann sum approximation to the)
desired contribution to the free energy, while the other terms are subdominant corrections. 
Optimizing over $ \ell $ we find $ \ell = \beta^{21/40} $ and we get the lower bound
$$f(\b) \geq C_0 \lf( \tx\frac12 \ri)^{-3/2} \beta^{-5/2} \left(1 - \mathcal O( \beta^{-\k })\right)$$
with $\k<1/40$. We now turn to the proof of Lemma 2.

\medskip

{\it Proof of Lemma 2.}
We first show that the eigenvalue equation implies the following remarkable inequality for $\r_E({\bf x}, {\bf y})$:
\be-\tilde \Delta\r_E({\bf x},{\bf y})\le 4E\rho_E({\bf x},{\bf y}),\label{s.3}\ee
where $\tilde \Delta$ is the Neumann Laplacian on the set $\{({\bf x},{\bf y}): {\bf x},\yy\in B,\ \xx\neq\yy\}$:
\bea -\tilde\Delta\r_E(\xx,\yy) &= & \disp\sum_{|\xx' - \xx|=1} \left[ \rho_E(\xx, \yy) \left( 1 - \delta_{\xx',\yy} \right) - \rho_E(\xx',\yy) \right]  \nonumber	\\
				& + & \disp\sum_{|\yy' - \yy|=1}  \left[ \rho_E(\xx, \yy)\left( 1 - \delta_{\yy',\xx} \right) - \rho_E(\xx,\yy') \right].	\nonumber
\eea 
To prove this, 
rewrite \eqref{s.2} as
$$  \tx\frac12  \disp\sum_{(\xx,\yy)} \lf[ a^\dagger_\xx (1-n_\yy) -a^\dagger_\yy (1-n_\xx) \ri] a_\xx (1-n_\yy), $$
where the sum is now over all {\em ordered} nearest neighbor pairs in $B$. Note that now the 
model looks like a system of hopping hard-core bosons, with the exclusion condition that they cannot hop on 
occupied sites and no additional interaction. A simple computation starting from $\langle E|H_{\L} a^\dagger_{\xx_1} a^\dagger_{\xx_2} a_{\xx_2} a_{\xx_1}
|E\rangle=E\langle E|a^\dagger_{\xx_1} a^\dagger_{\xx_2} a_{\xx_2} a_{\xx_1}|E\rangle$ 
shows that 
\bea &
E \rho_E(\xx_1,\xx_2)   =   \frac12 \disp\sum_{( \xx,\yy )  }  \Big\langle E\Big|  \lf[ a^\dagger_\xx (1-n_\yy) -a^\dagger_\yy (1-n_\xx) \ri] \nonumber\\
 &\times  \left( 
a^\dagger_{\xx_1} a^\dagger_{\xx_2} a_{\xx_2} a_{\xx_1}  + \delta_{\xx,\xx_1} n_{\xx_2} + \delta_{\xx,\xx_2} n_{\xx_1} \right) a_\xx (1 - n_\yy)  \, \Big| E \Big\rangle \,. \nonumber
\eea
The contribution of the first term $a^\dagger_{\xx_1} a^\dagger_{\xx_2} a_{\xx_2} a_{\xx_1}$ in the middle parenthesis is non-negative after summing over all pairs $(\xx,\yy)$, and can hence be dropped for a lower bound. For the remaining two 
terms, we rewrite $ a_\xx (1 - n_\yy)$  as 
\be \tx\frac 12 \left[  a_\xx (1 -n_\yy)  - a_\yy (1 -n_\xx) \right] +  \tx\frac 12 \left[  a_\xx (1 - n_\yy)  + a_\yy (1 - n_\xx) \right]
\nonumber\ee
and observe that the contribution of the first term yields again a non-negative expression. Hence  we get the lower 
bound 
\bea &
4E \rho_E(\xx_1,\xx_2)     \geq  \tx\frac14 \disp\sum_{( \xx,\yy )  }  \Big\langle E\Big|  \lf[ a^\dagger_\xx (1-n_\yy) -a^\dagger_\yy (1-n_\xx) \ri] \nonumber \\
 &\times  \left( \delta_{\xx,\xx_1} n_{\xx_2} + \delta_{\xx,\xx_2} n_{\xx_1} \right) \lf[ a_\xx (1 - n_\yy) + a_\yy (1 - n_\xx) \ri] \, \Big| E \Big\rangle \,. \nonumber
\eea
Elementary algebraic manipulations show that the right side is equal to $-\tilde\D\r_E(\xx_1,\xx_2)$, as desired.

We now explain how to infer Lemma 2 from \eqref{s.3}.
We extend $\rho_E$ to all of $\Z^3\times\Z^3$ by reflections
about the boundary of $B$ and by letting $\rho_E({\bf x},{\bf x})=0$ on the diagonal: more precisely, for $\vec m\in\mathbb Z^6$
we define the $\vec m$-th image point under reflections of a point $ \zv = (z_1, \ldots, z_6) \in B^2 $ as 
$$ z_j(m_j) = m_j \ell + \tx\frac 12 (\ell -1)  + (-1)^{m_j} \left( z_j - \tx\frac 12 (\ell -1) \right)  $$ 
and we let  $ \rho(\vec{z}(\vec{m})) \equiv \r_E(\vec{z}) $. This 
function satisfies for any $ \zv = ({\bf z}_1,{\bf z}_2) \in \Z^6 $ 
$$-\Delta \rho(\zv)\le 4E \rho(\zv)+2\rho(\zv)\chi^R(\zv)$$
where $\Delta$ is the lattice Laplacian on $\mathbb Z^6$ and $\chi^R(\zz_1,\zz_2)$ is equal to 1 if $\zz_1$ is at distance 1 from one of the images of $\zz_2$, and
0 otherwise. It plays the role of an interaction potential, which is non-local due to the reflections. 
The last inequality can equivalently be written as
$$\rho(\zv) \leq (1 - E/3)^{-1} \left( \langle \rho\rangle(\zv) + \tx\frac 1 {6}\r(\vec z) \chi^R(\zv) \right)$$
where $\langle \cdot\rangle(\zv)$ means averaging over nearest neighbors in $\mathbb Z^6$. In the last term on the right we bound $\r(\vec z)$ by 
$\|\r\|_\infty=\max_{\xx,\yy}\r(\xx,\yy)$. If we iterate $n$ times,
we further obtain ($*$ denoting the convolution)
\be \rho(\zv) \leq   (1 - E/3)^{-n} \bigg( (P_n*\rho)(\zv)+ \tx\frac 1 {6} \|\rho\|_\infty \disp\sum_{j=0}^{n-1}P_j*\chi^R(\zv) \bigg)\label{1.1}
\ee
where $P_n(\zv,\zv')$ denotes the probability that a simple symmetric random walk on $\mathbb Z^6$
starting at $\zv$ ends up at $\vec z'$ in $n$ steps. The idea used to derive a bound on $\lf\|\r\ri\|_\infty$ starting from 
\eqref{1.1} is most transparent in the slightly simplified case where $\c^R$ is replaced by $\c$, the characteristic function of the set $\{\zv=(\zz_1,\zz_2): |\zz_1-\zz_2|=1\}$. 
To treat the actual case, an additional argument is required \cite{CGS}, showing that the finite size of $B$ and the non-local part of the interaction $\c^R$
have a negligible effect on the magnitude  of $\r$.

We pick  $n\sim E^{-1}\gg 1$ in such a way that $(1 - E/3)^{-n}\simeq 1+\d$, with $\d$ a fixed small  constant.
From the central limit theorem, 
$$P_n(\zv,\zv')\simeq (3/(\pi n))^{3} e^{-3 |\zv-\zv'|^2/n}\;.$$
Therefore, $(P_n*\rho)(\zv)$ can be bounded from above by (const.)$E^3\sum_{\xx,\yy\in B}\r_E(\xx,\yy)$, which 
is smaller than (const.)$E^5\ell^4$, since the particle number is dominated by $E\ell^2$ thanks to Lemma 1. 
Moreover $\sum_{j=0}^{n-1}P_j(\zv,\zv')\le \sum_{j=0}^\infty P_j(\zv,\zv')=12\,G(\zv-\zv')$,
where $G$ is the Green's function of the Laplacian on $\mathbb Z^6$.
Therefore, replacing $\c^R$ by $\c$ in \eqref{1.1}, we find that the last term in \eqref{1.1}
is bounded by $2\lf\|\rho\ri\|_\infty G*\c(\zv)$. We have
$$\lf(G*\c \ri)(\zz_1,\zz_2)=\frac12\int e^{i\pp(\zz_1-\zz_2)}\frac{\sum_{i=1}^3\cos p_i}{\sum_{i=1}^3(1-\cos p_i)}\frac{\diff\pp}{(2\pi)^3}$$
which is smaller than 0.258, its value at $\zz_1=\zz_2$. Putting things together, we have shown that $\r(\vec z)$ is bounded by 
\be ({\rm const.})E^5\ell^4+2 \times 0.258 \times (1+\d) \lf\|\r \ri\|_\infty\;.
\nonumber\ee
If we choose $\d$ 
so small that $2 \times 0.258 \times (1+\d)<1$, Lemma 2 follows. 
$ \hfill\Box $

{\it Conclusions.} We report the first rigorous justification of the spin wave approximation for a quantum system with non-abelian continuous symmetry. 
We give precise bounds on the free energy at low temperatures, and establish spin order on suitable length scales. 

The research leading to these results has received funding from the European Research Council under the European Union's Seventh Framework Programme ERC Starting Grant CoMBoS (grant agreement 
n$^o$ 239694). M.C. acknowledges support from FIR grant ``Cond-Math" RBFR13WAET.



\end{document}